\documentclass[aps,preprint,nofootinbib,preprintnumbers]{revtex4}
\usepackage{graphicx,amsmath,amssymb,bm,subfigure,comment}

 \usepackage[english]{babel}
 \usepackage{color}

\newcommand{\beq}{\begin{equation}}
\newcommand{\eeq}{\end{equation}}
\newcommand{\Tr}{\mathrm{Tr}}

\begin{document}
\title{Nonrelativistic inverse square potential, \\ scale anomaly, and complex extension}
\author{Sergej Moroz}
\affiliation{Institut f\"{u}r Theoretische Physik Universit\"at Heidelberg Philosophenweg 16, D-69120 Heidelberg, Germany}
\author{Richard Schmidt}
\affiliation{Physik Department, Technische Universit\"at M\"unchen, James-Franck-Strasse, D-85748 Garching, Germany}
\date{9-2009}

\begin{abstract}
The old problem of a singular, inverse square potential in nonrelativistic quantum mechanics is treated employing a field-theoretic, functional renormalization method. An emergent contact coupling flows to a fixed point or develops a limit cycle depending on the discriminant of its quadratic beta function. We analyze the fixed points in both conformal and non-conformal phases and perform a natural extension of the renormalization group analysis to complex values of the contact coupling. Physical interpretation and motivation for this extension is the presence of an inelastic scattering channel in two-body collisions. We present a geometric description of the complex generalization by considering renormalization group flows on the Riemann sphere. Finally, using bosonization, we find an analytical solution of the extended renormalization group flow equations, constituting the main result of our work. 
\end{abstract}
\maketitle
\section{Introduction}
\label{introduction}
Exactly solvable problems play an important role in physics. They provide a backbone of our understanding and allow to develop perturbation theory on the basis of an exact solution. In nonrelativistic quantum mechanics the harmonic oscillator and the Coulomb central potential are certainly the two most important examples.
Another prominent example of an exactly solvable problem in quantum mechanics is the central, inverse square potential
\beq \label{i1}
V(r)=-\frac{\kappa}{r^2}.
\eeq
Most remarkably, in any spatial dimension the potential (\ref{i1}) is classically conformal invariant because it is a homogeneous function of degree -2 and has the same scaling as the nonrelativistic kinetic energy. Hence, the classical action is invariant under the nonrelativistic scaling $\vec{r}\to \rho \vec{r}$, $t\to \rho^2 t$.
The quantum problem with the potential (\ref{i1}) has a long history and is discussed both in standard textbooks \cite{FM, LandauQM} and in the scientific literature \cite{Case, Furlan}. More recently there were a number of renormalization group (RG) studies of this problem \cite{Kolomeisky,Gupta,Beane:2000wh,Barford:2002je,Bawin:2003dm,Ho,Braaten,Barford:2004fz,Hammer:2005sa,Hammer:2008ra,Kaplan}.

The quantum physics of the inverse square potential is well understood. The interaction (\ref{i1}) is an example of a singular potential \cite{singular} and must be treated with care. It is known that for the repulsive and weakly attractive coupling ($\kappa<\kappa_{cr}$)\footnote{$\kappa_{cr}$ represents the critical attractive coupling. In $d$ spatial dimensions it is given by $\kappa_{cr}=\frac{(d-2)^{2}}{4}$.}, the scale symmetry is preserved at the quantum level and the theory provides an example of nonrelativistic conformal field theory \cite{Nishida}. On the other hand, for strong attractions ($\kappa>\kappa_{cr}$) a discrete, geometric bound state spectrum develops in the two-body problem,  and the continuous scale symmetry is broken to a discrete subgroup by a quantum anomaly \cite{Camblong}. The anomaly has its origin in the singular short-distance behavior of the inverse square potential. In the RG language the anomaly manifests itself as a limit cycle instead of a scale-invariant fixed point.

The inverse square potential is a paradigmatic system for nonrelativistic conformal invariance and scale anomaly. Remarkably, there is a number of different physical systems, which are described (often only in some restricted domain) by the inverse square potential:
\begin{itemize}
\item The celebrated Efimov effect, first derived in \cite{Efimov}, consists in the formation of a tower of three-body bound states of identical non-relativistic bosons\footnote{More generally, for the occurance of the Efimov effect it is sufficient that at least two two-body subsystems have s-wave bound states close to the zero-energy threshold unless two of the three particles are identical fermions.} interacting through a short range potential. Exactly at resonance (unitarity regime) all scales drop out of the problem and the three-body spectrum becomes infinite and geometric. The three-body problem in quantum mechanics is treated most easily in hyperspherical coordinates \cite{NFJG}. 
Employing the adiabatic hyperspherical approximation \cite{Macek}, Faddeev decomposition of the wave function \cite{Faddeev} and restricting to the zero total angular momentum sector, one arrives at the remarkably simple one-dimensional effective equation \cite{BH}:
\beq \label{i3}
\left[-\frac{d^{2}}{d r^{2}}-\frac{s_{0}^{2}+1/4}{r^2} \right] \psi(r)=E \psi(r)
\eeq
with the Efimov parameter $s_0\approx 1.0062$.
This is a one-dimensional radial Schr\"odinger equation with the inverse square potential in the overcritical regime ($\kappa>\frac{1}{4}$).
\item The interaction of a polar molecule with an electron in three spatial dimensions can be approximated by a point dipole-charged particle anisotropic potential
\beq \label{i4}
V(\vec{r})\sim\frac{\cos{\theta}}{r^2},
\eeq
where the angle $\theta$ is measured with respect to the direction of the dipole moment. As was demonstrated in \cite{Camblong2}, this anisotropic conformal potential can be reduced to the effective isotropic inverse potential (\ref{i1}) in the zero angular-momentum channel. Remarkably, the dipole-electron system exhibits all interesting features characteristic for the potential (\ref{i1}). 
\item At the critical coupling $\kappa_{cr}$ the theory undergoes a transition from the conformal ($\kappa<\kappa_{cr}$) to the non-conformal ($\kappa>\kappa_{cr}$) regime. It has been demonstrated recently in \cite{Kaplan} that this transition is closely related to the classical Berezinskii-Kosterlitz-Thouless (BKT) phase transition in two dimensions. In particular, the energy $B$ of the lowest bound state near the critical coupling $\kappa_{cr}$ in the non-conformal regime vanishes like
\beq \label{i5}
B\sim \exp\left(-\frac{\pi}{\sqrt{\kappa-\kappa_{cr}}} \right)
\eeq 
which is analogous to the behavior of the inverse correlation length as the BKT transition temperature is approached from above \cite{Kosterlitz73}.
\item Quite unexpectedly, the inverse square potential arises in the near-horizon physics of some black holes. More specifically, this is the case for a massive, scalar field minimally coupled to gravity on the non-extremal spherically-symmetric Reissner-Nordstr\"om (RN) black hole background. It was demonstrated in \cite{Camblong} that the Klein-Gordon field equation of the scalar field reduces to the effective Schr\"odinger equation with the overcritical ($\kappa>\kappa_{cr}$) inverse square potential in the near-horizon limit. The related problem of a scalar particle near an extremal RN black hole was treated in \cite{Claus} leading to a similar finding.
\item Finally, the conformal inverse square potential appears naturally in the context of vacuum AdS/CFT correspondence \cite{Maldacena}. The field equation for a scalar field $\phi$ of mass $m$ in the Euclidean $AdS_{d+1}$ is
\beq \label{i6}
\partial_{r}^{2}\phi-\frac{d-1}{r}\partial_{r}\phi-\frac{m^2}{r^2}\phi-q^2\phi=0, \qquad q^2=(q^{0})^2+\vec{q}^2,
\eeq
where we transformed to the momentum space on the $AdS_{d+1}$ boundary $(x^{0},\vec{x})\to(q^{0},\vec{q})$, and $r$ denotes the radial direction in the $AdS_{d+1}$ space. We can change variable $\phi=r^{(d-1)/2}\psi$ and obtain
\beq \label{i7}
-\partial_{r}^{2}\psi+\frac{m^2+(d^2-1)/4}{r^2}\psi=-q^2\psi,
\eeq
which is a one-dimensional Schr\"odinger equation with an inverse square potential of strength $\kappa=-m^2-\frac{d^2-1}{4}$ and energy $E=-q^2$. As was emphasized in \cite{Kaplan}, the overcritical coupling $\kappa>\kappa_{cr}$ corresponds to the violation of the Breitenlohner-Freedman bound in $AdS_{d+1}$.
\end{itemize}
In this work we study the quantum problem of the inverse square potential using a functional renormalization method.
The plan of this paper is the following: In Sec. \ref{flowequation} we introduce our method, the physical system of interest, and derive the flow equations using the sharp cut-off regulator. A general mathematical discussion of the flow equation is performed in Sec. \ref{general}. Our central result is derived in Sec. \ref{extension}, where we extend the analysis to the complex plane, discuss the fixed point structure and find a numerical solution of the extended set of flow equations. Additionally, we provide a physical interpretation, geometric description and motivation for the complex extension. In Sec. \ref{bosonization} we bosonize the interaction. This allows us to view the problem from a different angle and, most remarkably, obtain an analytic solution of the generalized complex flow equations. We draw our conclusions in Sec. \ref{conclusion}. Finally, in Appendix \ref{appendix} we present the renormalization group flows on the Riemann sphere.

\section{The method, the model and the flow equation} 
\label{flowequation}
In this work we use the functional renormalization group and calculate a scale-dependent effective action functional $\Gamma_{k}$ \cite{Wetterich:1992yh} (for reviews see \cite{Berges:2000ew,Aoki:2000wm, Giesreview}) called also average action or flowing action. The method is formulated in Euclidean spacetime and employs the Matsubara formalism if one works at finite temperature. The flowing action $\Gamma_{k}$ includes all fluctuations with momenta $q\gtrsim k$. In the infrared (IR) limit $k\to 0$ the full quantum effective action $\Gamma=\Gamma_{k\to 0}$ is recovered and the problem is solved. In practice the dependence on the scale $k$ is introduced by adding a regulator term $R_{k}$ to the inverse propagator $\Gamma^{(2)}$. The flowing action $\Gamma_k$ obeys the exact functional flow equation \cite{Wetterich:1992yh}, and its bosonic version is given by
\begin{eqnarray}\label{fe}
  \partial_k \Gamma_k &=& \frac{1}{2} \Tr \,
  \partial_k R_k \, (\Gamma^{(2)}_k + R_k)^{-1} = \frac{1}{2} \Tr\,\tilde \partial_k \,\ln
  (\Gamma^{(2)}_k + R_k).
\end{eqnarray}
The functional differential equation for $\Gamma_{k}$ must be supplemented by an initial condition $\Gamma_{k\to\Lambda}=S$. The ``classical action'' $S$ describes the physics at the microscopic ultraviolet (UV) scale $k=\Lambda$ and is assumed to be known. In Eq. (\ref{fe}) $\Tr$ denotes a trace which sums over momenta, Matsubara frequencies, internal indices, and fields. The second functional derivative $\Gamma^{(2)}_{k}$ is the full inverse field propagator, which is modified by the presence of the IR regulator $R_k$.  The momentum dependent regulator function $R_{k}(q)$ must satisfy three important conditions, but otherwise can be chosen arbitrary \cite{Wetterich:1992yh}. The regulator dependence of the flowing action drops out for $k\to 0$, and one obtains the exact solution of Eq. (\ref{fe}). In the second form of the flow equation (\ref{fe}) $\tilde\partial_k$ denotes a scale derivative, which acts only on the IR regulator $R_{k}$. This form is convenient because it can be formulated in terms of one-loop Feynman diagrams. It is also convenient to introduce the RG ``time'' $t\equiv \ln(k/k_0)$, where $k_0$ is an arbitrary reference scale. The RG evolution towards the UV (IR) corresponds to $t>0$ ($t<0$). In the following we will use both $t$ and $k$.

In most cases of interest Eq. (\ref{fe}) can be solved only approximately. Usually one adopts some type of expansion of $\Gamma_{k}$ and then truncates at finite order. This leads to a finite system of ordinary differential equations. The expansions do not necessarily involve any small parameter and are generally of non-perturbative nature.  

Although our method allows us to address the quantum many-body problem at finite temperature, in this work we are interested only in the few-body (vacuum) physics, which is characterized by vanishing density ($n=0$) and vanishing temperature ($T=0$). In this case the effective action $\Gamma_{k=0}$, being the generating functional of the 1PI vertices, can be easily related to the different scattering amplitudes and bound state energies. Additionally, numerous simplifications arise in the structure of the flow equations (for a detailed discussion see \cite{DKS, MFSW}). 

Finally, we note that the three-body quantum problem for fermions and bosons was treated with functional renormalization in \cite{DKS, FSMW, MFSW, SFW}. In this work we apply functional renormalization to solve an arguably simpler problem, and our aim is twofold. First, the inverse square potential is a paradigm and it is important for understanding of other, more challenging problems. Second, working with this simple system, we develop a new renormalization group method of complexified flows in this paper. The method may be useful for more technical few-body problems in the future \cite{SM}.

In this work we study the nonrelativistic quantum mechanical problem of identical bosons interacting through a long-range potential (\ref{i1}) in $d$ spatial dimensions. The many-body field theory is defined in the UV by the microscopic action $S_{E}$
\beq \label{fe1}
\begin{split}
 S_{E}[\psi,\psi^{*}]&=\int_{0}^{1/T}d\tau \int d^{d}x \psi^{*}(\tau, \vec{x})[\partial_{\tau}-\Delta-\mu]\psi(\tau, \vec{x}) \\
 &-\frac{\lambda_{\psi}}{2}\int_{0}^{1/T} d\tau d^{d}x \psi^{*}(\tau, \vec{x})\psi^{*}(\tau, \vec{x})\psi(\tau, \vec{x})\psi(\tau, \vec{x}) \\
& -\int_{0}^{1/T} d\tau \int d^{d}x d^{d}y \psi^{*}(\tau, \vec{x})\psi^{*}(\tau, \vec{y})\frac{\kappa}{|\vec{x}-\vec{y}|^{2}}\psi(\tau, \vec{y})\psi(\tau, \vec{x}).
\end{split}
\eeq 
Our convention is $\hbar=2M_{\psi}=1$ with the boson mass $M_{\psi}$. We work in the Matsubara formalism with Euclidean time $\tau\in(0,1/T)$. In what follows we will be interested exclusively in the few-body (vacuum) physics. The vacuum state is characterized by zero density, which corresponds to zero chemical potential ($\mu=0$), and zero temperature ($T=0$). The bare action (\ref{fe1}) is invariant under a global $U(1)$ transformation and possesses Galilean space-time symmetry. The microscopic bare parameter $\kappa$ characterizes the strength of the long-range potential and is positive in the attractive case. We augmented the theory by a four-boson contact interaction term with a coupling $\lambda_{\psi}$. This is a consequence of the fact that the inverse square potential is singular at the origin \cite{Case, singular}. On its own the singular potential is not sufficient to define a quantum mechanical problem and must be augmented by the boundary condition at the origin \cite{Kaplan}. We will see that the introduction of $\lambda_{\psi}$ in the UV determines the $r=0$ boundary condition and makes the problem well-defined. 

\begin{figure}[t]
 \begin{center}
 \includegraphics[width=6.0in]{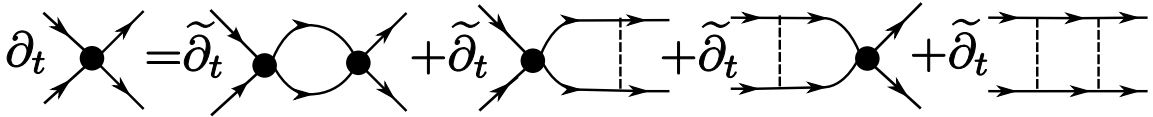}
 \end{center}
 \vskip -0.7cm \caption{Flow equation of the four-particle contact coupling $\lambda_{\psi}$ in form of Feynman diagrams. The solid lines with an arrow denote the boson $\psi$ regularized propagator, the dashed lines correspond to the long-range interaction vertex, and the dark blobs represent the contact coupling $\lambda_{\psi}$.}
\label{feyn}
 \end{figure}

For our method it is convenient to switch to momentum space. The Fourier transform of the $1/r^2$ potential in $d$ spatial dimensions reads
\beq \label{fe2}
\begin{split}
F_{d}(l)&=\int d^{d}r \frac{1}{r^2}\exp[i\vec{l}\cdot\vec{r}]=(2\pi)^{d/2}|\vec{l}|^{2-d}\int_{0}^{\infty}dz z^{d/2-2}J_{d/2-1}(z)= \\
& =\frac{(4\pi)^{d/2}\Gamma(d/2-1)|\vec{l}|^{2-d}}{4} \qquad 2<d<5.
\end{split}
\eeq
The restriction to dimensions $d$ in the range $2<d<5$ can be understood easily by the fact that on the one side, $d=2$ is a natural lower dimension, in which the integral (\ref{fe2}) is IR logarithmically divergent. On the other side, the upper bound $d=5$ can be relaxed, if we modify the Fourier integral by the introduction of a UV suppression factor $\exp(-\epsilon |\vec{r}|)$ and perform the limit $\epsilon\to 0$ in the very end. Hence, we will use
\beq \label{fe3}
F_{d}(l)=\frac{(4\pi)^{d/2}\Gamma(d/2-1)|\vec{l}|^{2-d}}{4} \qquad d>2.
\eeq
In what follows we consider only $d>2$.
First, we consider a momentum-independent (pointlike) truncation for the flowing action
\begin{equation} \label{fe3a}
\begin{split}
\Gamma_{k}[\psi,\psi^{*}]=&\int_{Q}\psi^{*}(Q)[i\omega+\vec{q}^2]\psi(Q)- \\
&-\kappa \int_{Q_1,Q_2,...,Q_4} F_{d}(l)\psi^{*}(Q_1)\psi(Q_2)\psi^{*}(Q_3)\psi(Q_4)\delta(-Q_1+Q_2-Q_3+Q_4) \\
& -\frac{\lambda_{\psi}}{2} \int_{Q_1,Q_2,...,Q_4} \psi^{*}(Q_1)\psi(Q_2)\psi^{*}(Q_3)\psi(Q_4)\delta(-Q_1+Q_2-Q_3+Q_4),
\end{split}
\end{equation}
where $Q=(\omega,\vec{q})$ and $\int_{Q}=\int\frac{d\omega}{2\pi}\int\frac{d^{d}q}{(2\pi)^{d}}$. The vector $\vec{l}=\vec{q}_2-\vec{q}_1=\vec{q}_3-\vec{q}_4$ gives the spatial momentum transfer during a collision and $l=|\vec{l}|$. In the nonrelativistic vacuum the propagator of the elementary field $\psi$ is not renormalized because the only diagram, which contributes to its flow, contains a hole (antiparticle) in the loop. As there are only particles but not holes in the nonrelativistic vacuum the propagator keeps its microscopic form. Remarkably, the coupling $\kappa$, characterizing the strength of the long-range $1/r^2$ potential, is also constant during the renormalization group flow (we discuss this issue in more detail in Sec. \ref{bosonization}). The only coupling which flows during the RG evolution in our truncation is the contact coupling $\lambda_{\psi}$. Its flow equation is
\beq \label{fe4}
\partial_{t}\lambda_{\psi}=\int_{L}\widetilde{\partial}_{t}\frac{[\lambda_{\psi}+2F_{d}(l)\kappa]^2}{(i\omega+l^2+R_{k}(L))(-i\omega+l^2+R_k(-L))}.
\eeq 
The flow equation is depicted in terms of Feynman diagrams in Fig. \ref{feyn}. 
In order to proceed further we must specify the cut-off function $R_k(L)$.

Nevertheless, we note that already at this point it is possible to identify the generic form of the flow equation which reads
\beq \label{fe5}
\partial_t \lambda_{\psi}=\alpha \lambda_{\psi}^{2}+\beta \lambda_{\psi}+\gamma
\eeq
with $\alpha,\beta,\gamma\in \mathbb{R}$, which depend on the coupling $\kappa$, dimension $d$ and a concrete choice of the cut-off function. We postpone the general analysis of equation (\ref{fe5}) until Sec. \ref{general}.

 It is convenient to rewrite Eq. (\ref{fe4}) as
\beq \label{fe6}
\partial_{t}\lambda_{\psi}=J_{0,d}\lambda_{\psi}^2+4 J_{1,d}\kappa \lambda_{\psi}+4J_{2,d} \kappa^2,
\eeq
where we defined the cut-off dependent integrals
\beq \label{fe7}
J_{n,d}\equiv \int \frac{d\omega}{2\pi}\frac{d^{d}l}{(2\pi)^d}\partial_t \frac{F^{n}_{d}(l)}{(i\omega+l^2+R_k(L))(-i\omega+l^2+R_k(-L))},
\eeq
where $F^{n}_{d}(l)\equiv F_{d}(l)^{n}$.
In the rest of this section we finish the computation by specifying the sharp regulator $R_{k}(L)=(i\omega+l^2)\left(\frac{1}{\theta(l^2-k^2)}-1 \right)$.

The cut-off function $R_{k}(L)=(i\omega+l^2)\left(\frac{1}{\theta(l^2-k^2)}-1 \right)$ cuts off quantum fluctuations sharply, i.e. it totally suppresses the modes with $l^2<k^2$ during the renormalization group evolution. This type of cut-off was used in the closely related three-body problem in \cite{MFSW}. The cut-off satisfies
\beq \label{sc1}
\frac{1}{i\omega+l^2+R_k(L)}=\theta(l^2-k^2)\frac{1}{i\omega+l^2}.
\eeq
Employing this property we can calculate $J_{n,d}$ defined in Eq. (\ref{fe7}) explicitly as
\beq \label{sc2}
J_{n,d}=\int \frac{d\omega}{2\pi}\frac{d^{d}l}{(2\pi)^d}\partial_t \theta(l^2-k^2) \frac{F^{n}_{d}(l)}{(i\omega+l^2)(-i\omega+l^2)}=-\frac{\pi S_d}{(2\pi)^{d+1}}k^{d-2}F_{d}^{n}(k)
\eeq
with $S_{d}=\frac{2\pi^{d/2}}{\Gamma(d/2)}$ being the area of a unit sphere in d-dimensional space. The resulting $J_{n,d}$ can be readily substituted in Eq. (\ref{fe6})
\beq \label{sc3}
\partial_t \lambda_{\psi}=-\frac{k^{d-2}}{(4\pi)^{d/2}\Gamma(d/2)}\lambda_{\psi}^{2}-\frac{2\kappa}{d-2}\lambda_{\psi}-\frac{(4\pi)^{d/2}\Gamma(d/2)}{(d-2)^2}\kappa^2.
\eeq
The contact coupling $\lambda_{\psi}$ has a naive (Gaussian) scaling dimension $[\lambda_{\psi}]=2-d$ and is IR irrelevant in $d>2$. Our aim is to investigate the fixed point structure of Eq. (\ref{sc3}). For this reason we introduce a dimensionless, rescaled coupling $\lambda_{\psi R}\equiv\frac{k^{d-2}\lambda_{\psi}}{(4\pi)^{d/2}\Gamma(d/2)}$ and its flow equation reads
\beq \label{sc4}
\partial_t \lambda_{\psi R}= -\lambda_{\psi R}^{2}+\left(-\frac{2\kappa}{d-2}+d-2 \right) \lambda_{\psi R}-\frac{\kappa^2}{(d-2)^2}.
\eeq
As will be shown in Sec. \ref{general}, the discriminant $D$ of the quadratic $\beta$-function determines the overall behavior of the solution. In our case $D=-4\kappa+(d-2)^2$. The critical $\kappa_{cr}$ is defined by the condition $D=0$ yielding
\beq \label{sc5}
\kappa_{cr}=\frac{(d-2)^2}{4},
\eeq   
which is in agreement with the quantum-mechanical non-perturbative calculation \cite{Kaplan}. For $\kappa<\kappa_{cr}$ (weak attraction and repulsion) the dimensionless coupling $\lambda_{\psi R}$ has two real fixed points
\beq \label{sc6}
\lambda_{\psi R}^{\pm}=-\frac{\kappa}{d-2}+\frac{d-2}{2}\pm\sqrt{\frac{(d-2)^2}{4}-\kappa}
\eeq
and the theory is scale invariant in the IR respectively UV. For $\kappa>\kappa_{cr}$ (strong attraction) the coupling $\lambda_{\psi R}$ ceases to have real fixed points and scale invariance is lost.
\section{General analysis of the flow equation} \label{general}
The flow equation for $\lambda_{\psi R}$ found in Sec. \ref{flowequation} has the general form
\beq \label{eft6}
\frac{d}{dt}\lambda_{\psi R}(t)=\alpha \lambda_{\psi R}(t)^2+\beta \lambda_{\psi R}(t)+\gamma,
\eeq
where $\alpha,\beta,\gamma\in\mathbb{R}$ are numerical coefficients. Without loss of generality we consider $\alpha\le 0$. This choice corresponds to the result obtained in Sec. \ref{flowequation}.
\begin{figure}[t]
 \begin{center}
 \includegraphics[width=3.0in]{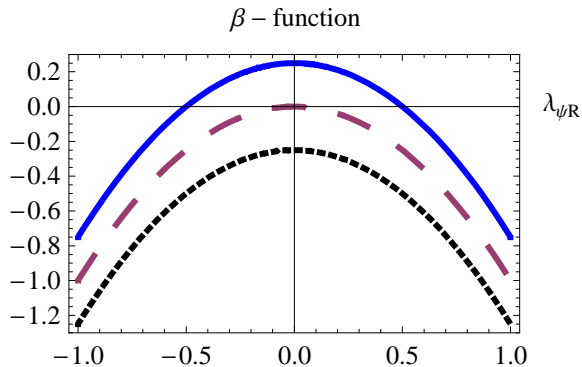} 
 \end{center}
 \vskip -0.7cm \caption{The $\beta$-function of the coupling $\lambda_{\psi R}$ for $\alpha=-1$ and $\beta=0$. The three lines correspond to different values of $\gamma$ with $D=1$ (solid blue), $D=0$ (dashed red) and $D=-1$ (dashed black).}
\label{figure2}
 \end{figure}
The form of the solution of Eq. (\ref{eft6}) is determined by the sign of the discriminant $D=\beta^2-4\alpha\gamma$ of the quadratic $\beta$-function. We consider the three different cases depicted in Fig. \ref{figure2}
\begin{itemize}
\item $D > 0$

In this case the $\beta$-function has two real fixed points $\lambda_{\psi R}^{IR}$ and $\lambda_{\psi R}^{UV}$ with $\lambda_{\psi R}^{IR}<\lambda_{\psi R}^{UV}$. The solution of Eq. (\ref{eft6}) depends on the interval, where the UV initial condition $\lambda_{\psi R}$ belongs. For the initial condition $\lambda_{\psi R}^{IR}<\lambda_{\psi R}<\lambda_{\psi R}^{UV}$ the solution is attracted towards the IR by the fixed point $\lambda_{\psi R}^{IR}$ and has the form:
\beq \label{eft7}
\lambda_{\psi R}(t)=\frac{-\beta-\sqrt{D}\tanh \left[\frac{\sqrt{D}}{2}(t+\eta)\right]}{2\alpha},
\eeq  
with $\eta$ fixed by the UV initial condition $\lambda_{\psi R}(t=0)$. For the initial condition $\lambda_{\psi R}>\lambda_{UV}$ the flow runs into a positive divergence (Landau pole), but reemerges at negative infinity and subsequently approaches $\lambda_{\psi R}^{IR}$ in the IR. The explicit solution in this case reads
\beq \label{eft7aa}
\lambda_{\psi R}(t)=\frac{-\beta+\sqrt{D}\coth \left[\frac{\sqrt{D}}{2}(-t+\eta)\right]}{2\alpha},
\eeq  
with $\eta$ fixed by the UV initial condition. In fact, the case $\kappa=0$, which leads to $D>0$, corresponds to the well-studied case of a contact interaction in atomic physics. In this context the appearance of the Landau pole signals the presence of a weakly bound molecular (dimer) state. Finally, for the initial condition $\lambda_{\psi R}<\lambda_{\psi R}^{IR}$ the RG flow is a smooth, monotonic function $\lambda_{\psi R}(t)$, which approaches $\lambda_{\psi R}^{IR}$ with the explicit solution given again by Eq. (\ref{eft7aa}). Notably, for $D>0$ the IR value $\lambda_{\psi R}(k=0)=\lambda_{\psi R}^{IR}$ is not sensitive to the concrete choice of the initial condition. We offer an elegant geometric description of the flow in Sec. \ref{extension} and Appendix \ref{appendix}.
\item $D=0$

This is a limit of the previous case ($D>0$) when the two fixed points merge. Excluding the trivial choices $\alpha=\beta=\gamma=0$ and $\alpha=\beta=0$; $\gamma\ne0$, the $\beta$-function has a single fixed point $\lambda_{\psi R}^{*}=-\frac{\beta}{2\alpha}$ (see Fig. \ref{figure2}).
 The RG equation takes the form
\beq \label{eft7a}
\frac{d \lambda_{\psi R}(t)}{dt}=\alpha[\lambda_{\psi R}(t)-\lambda_{\psi R}^{*}]^2
\eeq
with the solution
\beq \label{eft7b}
\lambda_{\psi R}(t)=\lambda_{\psi R}^{*}-\frac{1}{\alpha t +\eta},
\eeq
where $\eta$ is fixed by the initial condition. The running of the coupling $\lambda_{\psi R}(k)$ is logarithmic (which corresponds to the marginal deformation) and for $\eta<0$ it hits a IR Landau pole. The coupling runs into the pole at the scale $t=-\frac{\eta}{\alpha}$, nevertheless the RG evolution can be extended beyond this scale and approaches $\lambda_{\psi R}^{*}$ at $k=0$.
\item $D < 0$

In this case there are no real fixed points. The formal solution can be written as:
\beq \label{eft8}
\lambda_{\psi R}(t)=\frac{-\beta+\sqrt{-D}\tan \left[\frac{\sqrt{-D}}{2}(t+\eta)\right]}{2\alpha},
\eeq
where $\eta$ is fixed by the initial condition. This solution is periodic with a period $T=\frac{2\pi}{\sqrt{-D}}$. Remarkably, in this case the UV initial condition $\lambda_{\psi R}(k=\Lambda)$ plays an important role as it determines the infrared value $\lambda_{\psi R}(k=0)$.\footnote{This is analogous to the necessity to introduce the well-know three-body parameter in the context of the Efimov effect \cite{BH}.}
\end{itemize}
\begin{figure}[t]
 \begin{center}
 \includegraphics[width=3.0in]{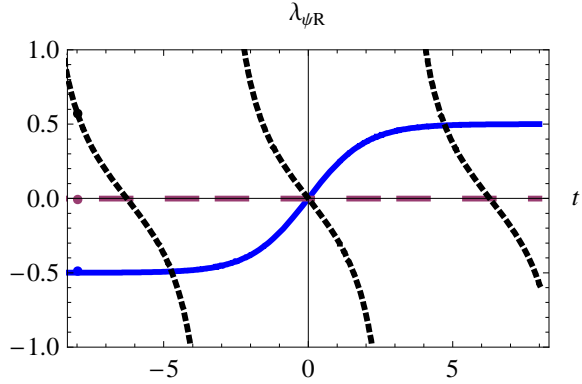}
 \end{center}
 \vskip -0.7cm \caption{The renormalization group flows of the coupling $\lambda_{\psi R}$ for $\alpha=-1$ and $\beta=0$ and initial condition $\lambda_{\psi R}(t=0)=0$. The different lines correspond to three different choices of $\gamma$ with $D=1, \eta=0$ (solid blue), $D=0, \eta\to\infty$ (dashed red), and $D=-1, \eta=0$ (dashed black).}
\label{figure3}
 \end{figure}
We plot the RG flows of $\lambda_{\psi R}$ in the three cases in Fig. \ref{figure3}. While for $D\ge 0$ the flows approach a fixed point in IR, the renormalization group evolution develops a limit cycle\footnote{Strictly speaking, to find a limit cycle one needs at least two couplings connected by the RG flow equations (see, for example, Eqs. (\ref{sb9}, \ref{sb9b})). In the case of a single coupling constant an infinite (unbounded) limit cycle appears only if there are periodic real discontinuities in the RG flow. We provide an elegant description of the infinite limit cycle on the Riemann sphere in Sec. \ref{extension}.} for $D<0$. As the RG scale $k$ can be related to the energy of the particles, the physical interpretation of the limit cycle solution is clear: During the RG flow one hits bound states, manifesting themselves as divergences of the coupling $\lambda_{\psi R}$. And since there are infinitely many divergences one has an infinite tower of bound states with a geometric spectrum for the case $D<0$.
\section{Complex extension} \label{extension}
In this section we discuss the different subcases, considered in Sec. \ref{general}, in more detail and extend the analysis to complex values of the interaction coupling. We also present a physical interpretation of this extension.
\subsection{Negative discriminant: complex fixed points} 
\label{complex}
\begin{figure}[t]
 \begin{center}
 \includegraphics[width=4.0in]{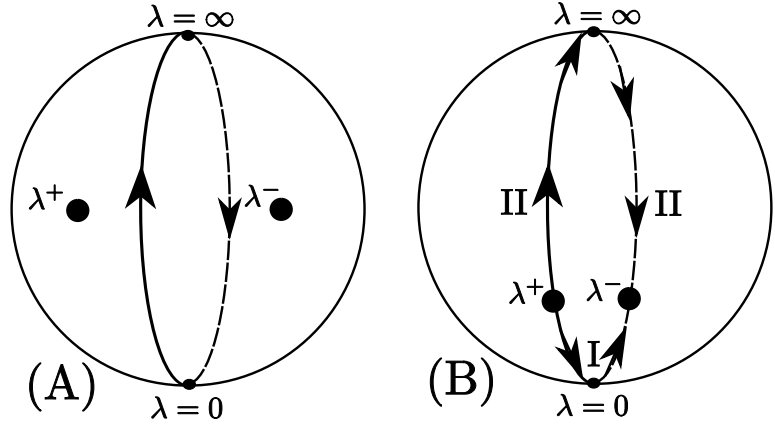}
 \end{center}
 \vskip -0.7cm \caption{The real flows of $\lambda$ projected onto the Riemann sphere: Great circle with arrows pointing towards the UV represents the real line. (A) For negative discriminant the fixed points $\lambda^+$ and $\lambda^-$ are complex, and the flow traverses the real great circle periodically generating a limit cycle. (B) For positive discriminant the IR fixed point $\lambda^+$ and the UV fixed point $\lambda^{-}$ lie on the real great circle. The UV fixed point can be reached from the IR fixed point via two different paths-- the ``short'' path I and the ``long'' path II.}
\label{riemannfig}
 \end{figure}

The renormalization group flow equation in our point-like approximation is\footnote{This equation coincides with Eq. (\ref{eft6}). To simplify notation, we denote the coupling $\lambda_{\psi R}$ as $\lambda$ in this section.}
\beq \label{c1a}
\partial_t \lambda=\alpha \lambda^2+\beta \lambda +\gamma
\eeq
with $\alpha, \beta, \gamma\in \mathbb{R}$, and one can vary the parameters $\beta$ and $\gamma$ by considering different $\kappa$ and $d$ in Eq. (\ref{sc4}). It is sometimes useful to express Eq. (\ref{c1a}) in the alternative form
\beq \label{c1b}
\partial_t \lambda=\alpha (\lambda-\lambda_{*})^2+\Delta\gamma, \qquad \lambda_{*}=-\frac{\beta}{2\alpha}, \qquad \Delta\gamma=\gamma-\frac{\beta^2}{4\alpha}.
\eeq
If the discriminant $D=\beta^2-4\alpha \gamma=-4\alpha\Delta\gamma<0$, the $\beta$-function has a pair of complex roots
\beq
\lambda^{\pm}=\frac{-\beta\pm i\sqrt{|D|}}{2\alpha}=\lambda_{*}\pm i \frac{\sqrt{|\alpha \Delta\gamma |}}{\alpha}.
\eeq
For this reason it is natural to consider the RG evolution of a generally complex variable $\lambda=\lambda_1+i \lambda_2$ in Eq. (\ref{c1a}). The resulting flow equations in the complex plane now read
\beq \label{c2}
\begin{split}
& \partial_t \lambda_1=\alpha \lambda_{1}^2-\alpha\lambda_{2}^2+\beta \lambda_1 +\gamma \\
& \partial_t \lambda_2=2\alpha \lambda_{1}\lambda_2+\beta \lambda_2.
\end{split}
\eeq
Notably, if we start on the real axis (i.e. $\lambda_2(t=0)=0$), the flow remains real
\beq
\begin{split}
& \partial_t \lambda_1=\alpha \lambda_{1}^2+\beta \lambda_1 +\gamma \\
& \partial_t \lambda_2=0
\end{split}
\eeq
with the periodic solution
\beq \label{c1}
\lambda_1(t)=\frac{-\beta+\sqrt{-D}\tan[\frac{\sqrt{-D}}{2}(t+\eta)]}{2\alpha},
\eeq
where $\eta$ is fixed by the initial condition $\lambda_1(t=0)$. The solution can be nicely identified with a limit cycle if we map the complex plane of the Riemann sphere. On the Riemann sphere the real line corresponds to  a great circle and the solution (\ref{c1}) traverses this circle periodically (see Fig. \ref{riemannfig} (A) and Appendix \ref{appendix}). Remarkably, on the complex plane the real axis is the separatrix of the two complex fixed points.

Let us investigate the properties of the fixed points $\lambda^{\pm}$ by considering the stability matrices
\beq \label{c3}
M^{\pm}_{ij}=\frac{\partial \beta_i}{\partial\lambda_j}|_{\lambda^{\pm}}.
\eeq
A straightforward computation shows
\beq \label{c3a}
M^{\pm}=\left( \begin{array}{cc}
0 & -b^{\pm} \\
b^{\pm} & 0
\end{array} \right),  \qquad b^{\pm}=\pm i\sqrt{|D|}.
\eeq
The stability matrices $M^{\pm}$ have a pair of complex conjugate pure imaginary eigenvalues
\beq
\kappa_{1}^{\pm}=\pm i\sqrt{|D|} \qquad \kappa_{2}^{\pm}=(\kappa_{1}^{\pm})^{*}.
\eeq 

Therefore, the local flow near the fixed points has a form of a circle. The sign of $b^{\pm}$ in the matrix (\ref{c3a}) determines the orientation of the circle. For $b^{\pm}>0$ the circulation is anticlockwise, while for $b^{\pm}<0$ it is clockwise. Hence the fixed points $\lambda^{\pm}$ have the opposite circulation.

\begin{figure}
\centering
	\includegraphics[height=2.5in]{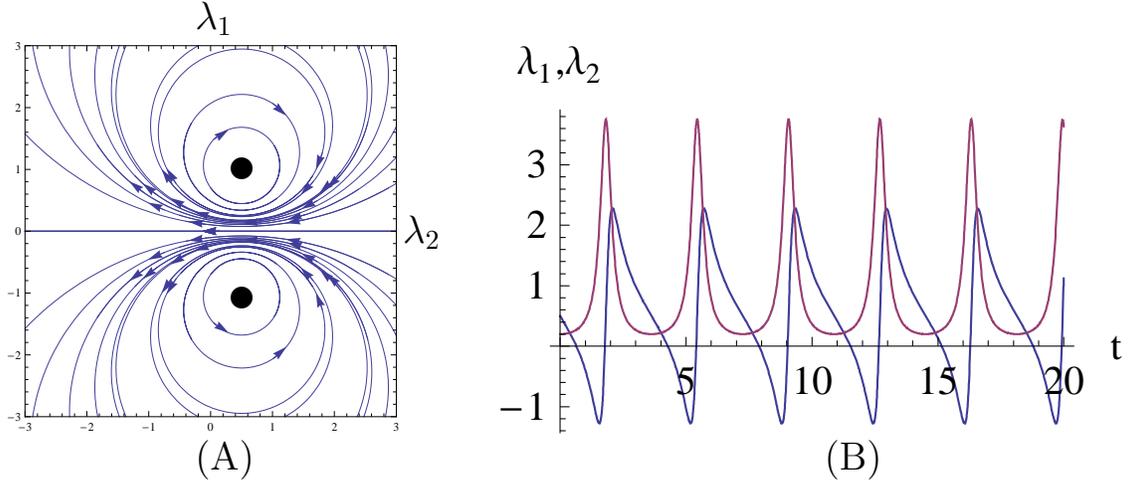}
	\caption{(A) The phase portrait of the flow equations (\ref{c2}) for the specific choice of parameters $\alpha=\gamma=-1$ and $\beta=1$. Arrows denote the direction towards the UV. (B) Corresponding periodic flows of the real part $\lambda_1$ (blue) and the imaginary part $\lambda_2$ (red) of the complex coupling $\lambda$.}
\label{fig1}
\end{figure}

The phase portrait of the RG flow in the complex plane, computed numerically for specific values of parameters, and the example of the RG flow are depicted in Fig. \ref{fig1}.
We observe that the periodic divergences of the pure real solution (\ref{c1}) are regularized and the flow of the imaginary part $\lambda_2$ develops a tower of resonances. The analytic solution of Eq. (\ref{c2}) for $D<0$ is given in Sec. \ref{bosonization}.
\subsection{Positive discriminant: real fixed points} \label{real}
\begin{figure}
\centering
	\includegraphics[height=2.5in]{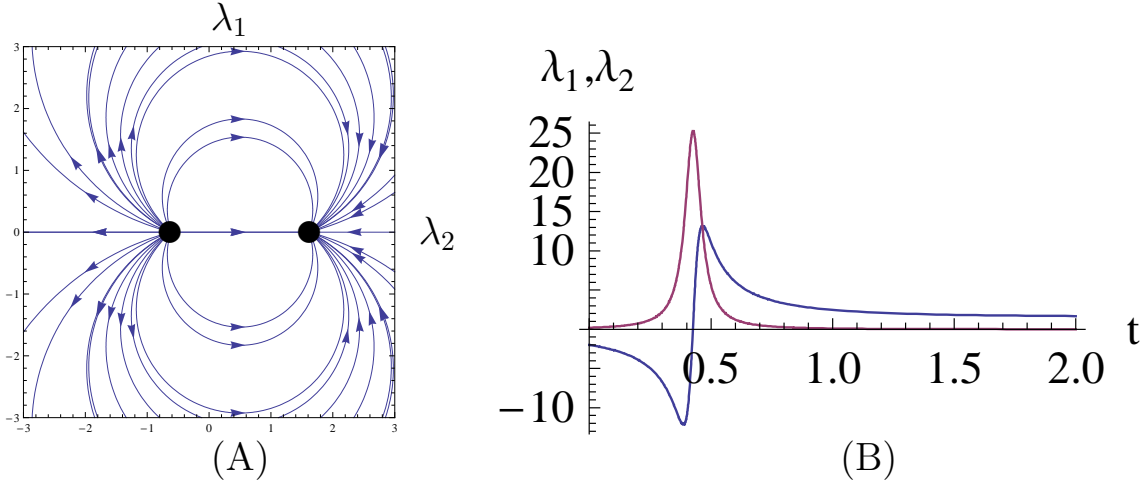}
	\caption{(A) The phase portrait of the flow equations (\ref{c2}) for the specific choice of parameters $\alpha=-1$ and $\beta=\gamma=1$. Arrows denote the direction towards the UV. (B) Corresponding renormalization group flows of the real part $\lambda_1$ (blue) and the imaginary part $\lambda_2$ (red) of the complex coupling $\lambda$.}
\label{fig2}
\end{figure}

If the discriminant $D$ is positive, the $\beta$-function has a pair of real roots
\beq \label{rfp1}
\lambda^{\pm}=\frac{-\beta\pm \sqrt{D}}{2\alpha}.
\eeq
As before, we generalize the coupling to complex values $\lambda=\lambda_1+i\lambda_2$ and obtain the pair of coupled differential equations (\ref{c2}). The phase portrait and a specific solution of the flow equations (\ref{c2}) for $D>0$ are depicted in Fig. \ref{fig2}. We find the analytic solution of Eq. (\ref{c2}) for $D>0$ in Sec. \ref{bosonization}.

The character of the fixed points $\lambda_{\pm}$ can be determined from the stability matrices $M_{ij}^{\pm}$, defined by Eq. (\ref{c3}). For $D>0$, we obtain
\beq
M^{\pm}=\left( \begin{array}{cc}
\pm\sqrt{D} & 0 \\
0 & \pm\sqrt{D}
\end{array} \right)=\pm\sqrt{D}I 
\eeq
with the degenerate eigenvalues $\kappa^{\pm}$
\beq
\kappa^{\pm}\equiv \kappa_{1}^{\pm}=\kappa_{2}^{\pm}=\pm\sqrt{D}.
\eeq
The sign of the real eigenvalue determines whether the fixed point is UV attractive or repulsive. The eigenvalue $\kappa^{+}$ is positive and hence the fixed point $\lambda^{+}$ (left fixed point in Fig. \ref{fig2}(A)) is UV repulsive, meaning that as the sliding scale $k$ is increased the flow is driven away from $\lambda^{+}$. On the other hand $\kappa^{-}$ is negative and therefore the fixed point $\lambda^{-}$ (right fixed point in Fig. \ref{fig2}(A)) is UV attractive. Notice that any two-component vector is an eigenvector of the fixed points $\lambda^{\pm}$.

We propose a geometric interpretation for the different behavior of the real solution in dependence on the initial conditions, which we observed in Sec. \ref{general}. For a positive discriminant $D$ both the UV fixed point $\lambda^-$ and the IR fixed point $\lambda^+$ are situated on the large real circle on the Riemann sphere (see Fig. \ref{riemannfig} (B) and Appendix \ref{appendix}). We note that the UV fixed point can be reached from the IR fixed point following two different paths. Depending on the initial conditions $\lambda_{\text{in}}=\lambda^{+}\pm \epsilon$, the flow can follow either a ``short'' path I or a ``long'' path II (see Fig. \ref{riemannfig} (B)). Path I corresponds to the solution (\ref{eft7}), which is a standard way how to regularize the inverse square potential at $\kappa<\kappa_{cr}$. On the other hand, the path II traverses $\lambda=\infty$ on the Riemann sphere and corresponds to the solution (\ref{eft7aa}) with the Landau pole. The divergence during the RG evolution is identified with a single bound state, which, in the case of $\kappa=0$, becomes the the well-known shallow dimer studied extensively  in atomic physics \cite{BH}. We notice that on the full Riemann sphere the path II can be continuously deformed into the path I. Thus, the introduction of a small imaginary initial condition $\lambda_2(t=0)$ for the complex extended flow, which leads to a small deformation of the path II, regularizes the divergence in the real part of the coupling (see Fig. \ref{fig2} and Appendix \ref{appendix}).
\subsection{Zero discriminant: degenerate fixed points} \label{degen}
Finally, for $D=0$, the pair of roots of the $\beta$-function becomes degenerate
\beq
\lambda_{*}\equiv\lambda^{+}=\lambda^{-}=-\frac{\beta}{2\alpha}
\eeq
which corresponds to a single, real-valued fixed point $\lambda_{*}$.

\begin{figure}
\centering
	\includegraphics[height=2.5in]{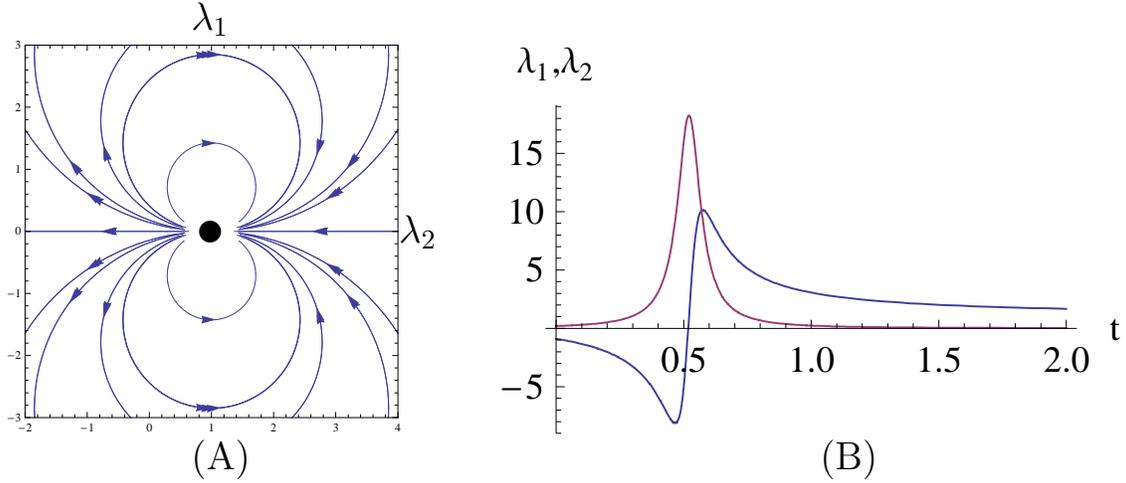}
	\caption{(A) The phase portrait of the flow equations (\ref{c2}) for the specific choice of parameters $\alpha=-1$, $\beta=2$ and $\gamma=-1$. Arrows denote the direction towards the UV. (B) Corresponding renormalization group flows of the real part $\lambda_1$ (blue) and the imaginary part $\lambda_2$ (red) of the complex coupling $\lambda$.}
\label{fig3}
\end{figure}

As it turns out, the stability matrix $M_{ij}=\frac{\partial \beta_{i}}{\partial \lambda_j}|_{\lambda_{*}}$ vanishes in the case of $D=0$. This corresponds to a logarithmic (marginal) renormalization group flow in the vicinity of the fixed point and local properties of the degenerate fixed point $\lambda_{*}$ are accounted for by the second derivative of the $\beta$-function $K_{ijl}$ at $\lambda_{*}$
\beq
K_{ijl}=\frac{\partial^{2}\beta_{i}}{\partial\lambda_{j}\partial\lambda_{l}}|_{\lambda^{*}} \qquad i,j,l=1,2.
\eeq
The only nontrivial components are
\beq
K_{111}=-K_{122}=K_{212}=K_{221}=2\alpha.
\eeq
Again, we computed the phase portrait and a specific solution of the flow equations (\ref{c2}) and show the results in Fig. \ref{fig3}. We observe that the Landau pole of the real part of $\lambda$ is regularized and the imaginary part of $\lambda$ develops a single resonance.
\subsection{Physical interpretation of the complex extension} \label{physint}
One may wonder about the physical meaning of the complex contact coupling $\lambda$ introduced in this section. To answer this question, we consider inelastic scattering\footnote{By definition inelastic collisions change the internal state of the colliding particles and, hence, at least one of the colliding particles must be composite and have some internal structure. For example, in experiments two atoms might collide and fall into energetically deeper lying internal states. The excess in energy will be converted into kinetic energy and the atoms will be lost if the released energy suffices to overcome the trapping potential.} of particles in a central potential in quantum mechanics. Following \cite{LandauQM} in three spatial dimensions the scattering amplitude in the center of mass frame can be expanded in partial waves
\beq \label{pi0}
f_k(\theta)=\sum_{l=0}^{l=\infty}(2l+1)f_l(k) P_l(\cos(\theta)),
\eeq
where $P_l(\cos\theta)$ are the Legendre polynomials. From the unitarity constraints the partial waves scattering amplitudes $f_l(k)$ as functions of scattering momentum $k$ are given by
\beq \label{pi1}
f_l(k)=\frac{1}{g_l(k^2)-ik}.
\eeq
Here, $g_l(k^2)$ is an even function of $k$ which is generally complex in the case of presence of inelastic channels.

For low-energy scattering of particles, interacting through a short-range potential, only the s-wave contribution $f=f_0$ is substantial and $g(k^2)=g_0(k^2)$ can be expanded as
\beq \label{pi2}
g(k^2)=-a^{-1}+\frac{1}{2}r_{\text{eff}} k^2+\dots,
\eeq
where in the case of elastic scattering $a$ and $r_{\text{eff}}$ are real and denote the scattering length and effective range.
However, in the inelastic case $a$ may also be regarded as a complex scattering length $a=\alpha+i\beta$ and the effective range $r_{\text{eff}}$ might also be complex \cite{Hussein}. In the forthcoming argument it is sufficient to consider low energy scattering and therefore we keep only the scattering length $a$ and neglect the second term in Eq. (\ref{pi2}).

The generalized optical theorem for the scattering of indistinguishable particles
\beq \label{pi3}
\text{Im} f(k)=\frac{k}{8\pi}\sigma_{tot}=\frac{k}{8\pi}(\sigma_{el}+\sigma_{in})
\eeq
holds also in the case of general inelastic scattering \cite{LandauQM}. In this case the total scattering cross section $\sigma_{tot}$ is the sum of the positive elastic $\sigma_{el}$ and inelastic $\sigma_{in}$ contributions. The positivity of $\sigma_{in}$ and the optical theorem (\ref{pi3}) imply that the imaginary part of the scattering length $\beta$ must be negative. At low energies and for sufficiently short-range interactions the complex scattering length fully determines the elastic and inelastic cross sections, and for indistinguishable particles the result reads \cite{LandauQM}
\beq \label{pi4}
\begin{split}
&\sigma_{el}=8\pi |a|^2(1-2k|\beta|) \\
&\sigma_{in}=8\pi \frac{|\beta|}{k}(1-2k|\beta|).
\end{split}
\eeq
We observe that a non-vanishing imaginary part of the scattering length $\beta$ is required to obtain an inelastic contribution to the total cross section\footnote{We note that according to \cite{LandauQM} Eq. (\ref{pi4}) is valid for sufficiently fast decrease of the interaction potential at large distances (at least $1/r^3$). Thus, it is not strictly applicable in our case of the long-range $1/r^2$ potential, and the presented argument has to be taken as heuristic.}, which results in loss of particles.

The presented arguments can be connected to a complex generalization of $\lambda$ by the simple observation that in a system of identical bosons with short-range interactions the physical scattering length $a$ is related to the IR value of the dimensionful, contact coupling $\lambda_{\psi}$ in $d=3$ via the simple formula (in the case of $\kappa=0$)
\beq \label{pi5}
\lambda_{\psi}(k=0)=-8\pi a.
\eeq
Thus, allowing for complex values of $\lambda_{\psi}$ during the RG flow corresponds to the presence of inelastic two-body collisions leading to particle loss. This provides a physical interpretation of the complex coupling $\lambda_{\psi}$. We must stress, however, that the local $(\psi^{\dagger}\psi)^2$ operator with a negative imaginary coefficient can describe inelastic scattering only to deep energetic states, i.e. the gap energy $E_{gap}$ of the state must be large compared to other energy scales in the problem \cite{BKP}. On the other hand, decay into shallow bound states can be described by introduction of a non-local operator with a complex coefficient. In fact, a large imaginary part of the two-body interaction might also be useful in the context of atomic physics. This case has recently been studied for cold atoms confined to motion in 1D. In particular it has been shown that a Tonks-Girardeau gas, where 1D strongly repulsive bosons exhibit fermionic like behavior, can also be induced by the effect of strong dissipation \cite{Syassen,Kiffner}, meaning large imaginary two-body interaction $\lambda_{\psi 2}$.

Finally, we note that our extension of the RG equation to the complex plane can be used as an efficient numerical tool. As was demonstrated in Sec. \ref{general}, the flow equation (\ref{eft6}) has the periodic solution (\ref{eft8}) in the non-conformal phase. The running coupling $\lambda_{\psi R}$ diverges periodically during the RG evolution, making the numerical solution impossible beyond the first divergence.  We showed in this section that adding a small imaginary part to the initial condition $\lambda_{\psi R}(t=0)$ makes the solution of Eq. (\ref{eft6}) numerically feasible. Physically, this corresponds to converting stable bound states to long-lived resonances. As shown in the next section, the problem considered in this work can be treated analytically and hence the numerical treatment is not necessary. Nevertheless, the extension turns out to be useful in some other few-body problems, which can not be solved analytically \cite{SM}.
\section{Bosonization} \label{bosonization}
In this section we approximate the four-particle  vertex by an exchange of a composite particle. In the literature \cite{Giesreview} this is known as bosonization procedure and it is a powerful, physically transparent concept. In this work we will bosonize in two distinct ways. The first bosonization corresponds to the exchange of a massless non-dynamical particle in the t-channel. In this realization the composite particle mediates the long-range $1/r^2$ interaction combined with the contact four-particle interaction. Alternatively, one can approximate the contact vertex by the exchange of a massive particle in the s-channel. While at low energies and momenta bosonization simply reproduces the pointlike (momentum-independent) approximation of Sec. \ref{flowequation},  it resolves some of momentum structure of the interaction at higher energies.

\subsection{t-channel bosonization} \label{tbos}
In this subsection we approximate the total four-particle interaction vertex by the exchange of a composite massless particle $\chi$ in the t-channel.
 \begin{center}
 \includegraphics[width=2.0in]{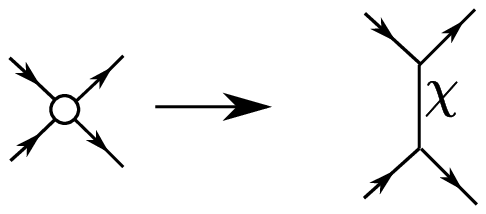}
 \end{center}
Galilean symmetry of the nonrelativistic vacuum implies that the inverse propagator $P_{\chi}(\omega,\vec{q})$ of the massless particle depends only on the absolute value of the spatial momentum $\vec{q}$, but not on the frequency $\omega$. Thus, the composite $\chi$ is not a dynamical particle and serves only to mediate the two-body long-range interaction.

Mathematically, bosonization is implemented by means of the Hubbard-Stratonovich transformation. At $T=\mu=0$ one adds  an auxiliary part to the microscopic action (\ref{fe1})
\beq \label{tb1}
\begin{split}
S_{E,aux}[\psi,\psi^{*},\chi]=\frac{1}{2}\int_{Q_1,Q_2,...,Q_4}P_{\chi}(l)&\{\chi(-L)+\frac{g}{2P_{\chi}(\vec{l})}\psi^{*}(Q_1)\psi(Q_2) \} \times \\  
& \{\chi(L)+\frac{g}{2P_{\chi}(\vec{l})}\psi^{*}(Q_3)\psi(Q_4) \} \times \\
& \delta(-Q_1+Q_2-Q_3+Q_4),
\end{split}
\eeq 
where $L=(\omega_{l},\vec{l})=Q_2-Q_1=Q_3-Q_4$ is the momentum transfer $d+1$-vector and $\chi$ is a real-valued field. The modified action $S_{E}^{'}=S_E+S_{E,aux}$ defines an equivalent quantum mechanical problem as the action $S_E$ because the functional integral in the field $\chi$ is Gaussian.

Notably, at the UV scale we can achieve the cancellation of the four-particle interaction term by choosing the non-local inverse propagator
\beq \label{tb2}
P_{\chi}(l)=\frac{g^2}{4}\left[\lambda_{\psi}^{UV}+2\kappa F_{d}(l)\right]^{-1},
\eeq
where $\lambda_{\psi}^{UV}$ is the microscopic (bare) contact coupling in Eq. (\ref{fe1}).
The modified bare action $S_{E}^{'}$ now reads
\beq \label{tb3}
\begin{split}
S_{E}^{'}[\psi,\psi^{*},\chi]=&\int_{Q}\psi^{*}(Q)[i\omega+\vec{q}^2]\psi(Q)+\frac{1}{2}\int_{Q}\chi(-Q)P_{\chi}(q)\chi(Q) \\
&+g\int_{Q_1,Q_2,Q_3}\psi^{*}(Q_1)\psi(Q_2)\chi(Q_3)\delta(-Q_1+Q_2+Q_3)
\end{split}
\eeq
with the four-boson interaction replaced by the Yukawa-like term $\chi\psi\psi^{*}$. Our truncation for the flowing action is chosen to be
\beq \label{tb4}
\begin{split}
\Gamma_k[\psi,\psi^{*},\chi]=&\int_{Q}\psi^{*}(Q)P_{\psi}(Q)\psi(Q)+\frac{1}{2}\int_{Q}\chi(-Q)P_{\chi}(Q)\chi(Q) \\
&+g\int_{Q_1,Q_2,Q_3}\psi^{*}(Q_1)\psi(Q_2)\chi(Q_3)\delta(-Q_1+Q_2+Q_3) \\
&-\frac{\lambda}{2}\int_{Q_1,Q_2,...,Q_4} \psi^{*}(Q_1)\psi(Q_2)\psi^{*}(Q_3)\psi(Q_4)\delta(-Q_1+Q_2-Q_3+Q_4),
\end{split}
\eeq
where $\lambda$ is a contact coupling, which is zero in the ultraviolet by construction and is regenerated during the RG flow through a box diagram.

Due to the numerous simplifications of the nonrelativistic vacuum, which are described in detail in \cite{MFSW}, we note
\begin{itemize}
\item the inverse propagators $P_{\psi}$ and $P_{\chi}$ are not renormalized during the RG flow
\beq \label{tb5}
P_{\psi}(Q)=i\omega+\vec{q}, \qquad P_{\chi}(Q)=\frac{g^2}{4}\left[ \lambda_{\psi}^{UV}+2\kappa F_{d}(l)\right]^{-1}.
\eeq
Technically this arises from the fact that the one-loop Feynman diagrams, which would renormalize $P_{\psi}$ and $P_{\phi}$, have poles in the same half plane of the complex loop frequency. One can close the integration contour such that it does not enclose any frequency poles. As the result of the residue theorem, the frequency integrals vanish, leading to the non-renormalization of the propagators. We observe the clear manifestation of the non-renormalization of the long-range potential, mentioned already in Sec. \ref{flowequation}. Another example for this behavior is the boundary sine-Gordon theory, which was studied in \cite{FZ86,Guinea85}.
\item The coupling $g$ can be chosen arbitrarily, e.g. $g=1$, at the UV scale. This is a well-know ambiguity of the Hubbard-Stratonovich decoupling. Furthermore, the coupling $g$ is also not renormalized during the renormalization group evolution for the reason mentioned in the previous point.
\item Interaction terms of the form $\chi^{n}$ with $n\ge 1$ and $\psi^{*}\psi \chi^{k}$ with $k>1$ are absent in the ultraviolet and are not generated during the flow due to the argument provided above. On the other hand interaction terms of the form $(\psi^{*}\psi)^{2}\chi^{k}$ with $k\ge 0$ are generated through box diagrams during the renormalization group evolution.
\item Our truncation is complete up to the two-body sector. This is a result of the special hierarchy of the flow equations in the nonrelativistic vacuum \cite{MFSW, DKS}, where the couplings from the higher-body sectors do not influence the couplings from the  lower-body sectors.
\end{itemize}

It is straightforward to derive the flow equation of the momentum-independent coupling $\lambda$. The result turns out to be given by Eqs. (\ref{fe6}, \ref{fe7}) with the substitution $F_{d}^{n}(l)\to F_{d}^{n}(l)+\frac{\lambda_{\psi}^{UV}}{2\kappa}$ in Eq. (\ref{fe7}). The initial condition $\lambda_{\psi}(k=\Lambda)=\lambda_{\psi}^{UV}$ of the one-channel model of Sec. \ref{flowequation} is implemented directly as the initial condition of the inverse propagator $P_{\chi}$ in the bosonization approach. Otherwise, the flows are completely equivalent at the level of our approximation.
\subsection{s-channel bosonization} \label{sbos}
\begin{figure}[t]
 \begin{center}
 \includegraphics[width=6.0in]{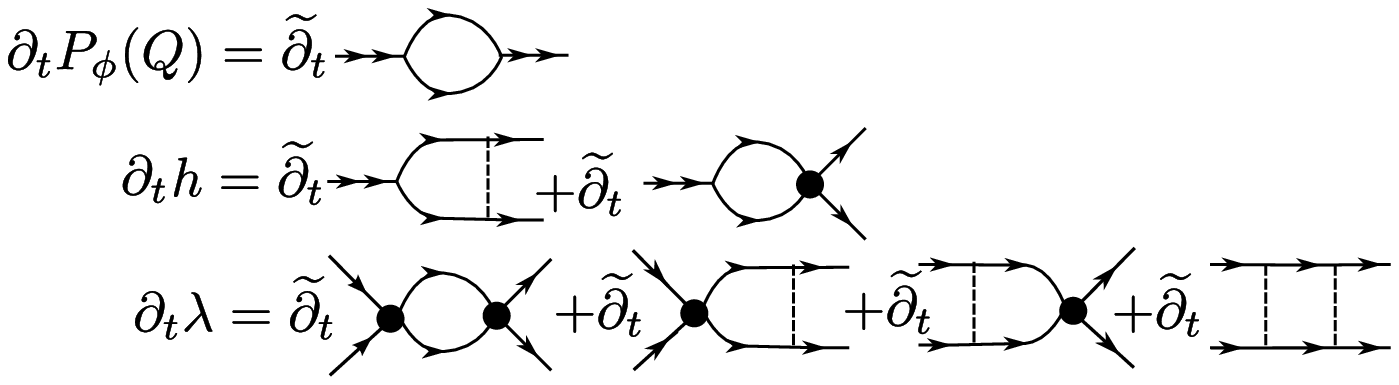}
 \end{center}
 \vskip -0.7cm \caption{Flow equations of $P_{\phi}(Q)$, $h$ and $\lambda$ in the form of Feynman diagrams. The solid lines with one (two) arrow(s) denote regularized propagator for the field $\psi$ ($\phi$), the dashed lines correspond to the long-range interaction vertex, the dark blob represents the contact coupling $\lambda$.}
\label{feyn1}
 \end{figure}

We will follow an alternative bosonization procedure in this subsection. In this approach the four-particle contact vertex is approximated by the exchange of a massive ($M_{\phi}=2M_{\psi}$) particle $\phi$ in the s-channel.
 \begin{center}
 \includegraphics[width=2.0in]{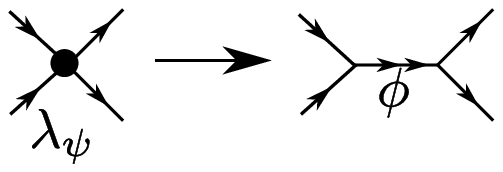}
 \end{center}
At the level of the microscopic (bare) action this is achieved via the Hubbard-Stratonovich transformation. The auxiliary part, added to the microscopic action (\ref{fe1}) at $T=\mu=0$, reads
\beq \label{sb1}
\begin{split}
S_{E,aux}[\psi,\psi^{*},\phi,\phi^{*}]=\int_{Q_1,Q_2,...,Q_4}m^2&\{\phi^{*}(Q_1+Q_3)+\frac{h}{2m^2}\psi^{*}(Q_1)\psi^{*}(Q_3) \} \times \\  
& \{\phi(Q_2+Q_4)+\frac{h}{2m^2}\psi(Q_2)\psi(Q_4) \} \times \\
& \delta(-Q_1+Q_2-Q_3+Q_4).
\end{split}
\eeq 
For the specific choice $\frac{\lambda_{\psi}}{2}=\frac{h^2}{4m^2}$, we can cancel the contact term in the modified microscopic action $S_{E}^{'}=S_{E}+S_{E,aux}$
\beq \label{sb2}
\begin{split}
S_{E}^{'}[\psi,\psi^{*},\phi,\phi^{*}]=&\int_{Q}\psi^{*}(Q)[i\omega+\vec{q}^2]\psi(Q)+\int_{Q}\phi^{*}(Q)m^2\phi(Q) \\
&+\frac{h}{2}\int_{Q_1,Q_2,Q_3}\left[\phi^{*}(Q_1)\psi(Q_2)\psi(Q_3)+ \phi(Q_1)\psi^{*}(Q_2)\psi^{*}(Q_3) \right]\delta(-Q_1+Q_2+Q_3) \\
&-\kappa \int_{Q_1,Q_2,...,Q_4} F_{d}(l)\psi^{*}(Q_1)\psi(Q_2)\psi^{*}(Q_3)\psi(Q_4)\delta(-Q_1+Q_2-Q_3+Q_4).
\end{split}
\eeq
During the renormalization group evolution the massive field $\phi$ becomes dynamical and develops an inverse propagator $P_{\phi}(Q)=f(i\omega+\frac{\vec{q}^2}{2}+m^2)$, where $f$ is some yet unknown function. The argument of the inverse propagator is fixed by the Galilean symmetry of the nonrelativistic vacuum, while the function $f$ is determined by the dynamics. Our truncation of the flowing action within the s-channel bosonization approach is
\beq \label{sb3}
\begin{split}
\Gamma_k[\psi,\psi^{*},\phi, \phi^{*}]=&\int_{Q}\psi^{*}(Q)P_{\psi}(Q)\psi(Q)+\int_{Q}\phi^{*}(Q)P_{\phi}(Q)\phi(Q) \\
&+\frac{h}{2}\int_{Q_1,Q_2,Q_3}\left(\phi^{*}(Q_1)\psi(Q_2)\psi(Q_3)+ \phi(Q_1)\psi^{*}(Q_2)\psi^{*}(Q_3) \right)\delta(-Q_1+Q_2+Q_3) \\
&-\kappa \int_{Q_1,Q_2,...,Q_4} F_{d}(l)\psi^{*}(Q_1)\psi(Q_2)\psi^{*}(Q_3)\psi(Q_4)\delta(-Q_1+Q_2-Q_3+Q_4) \\
&-\frac{\lambda}{2}\int_{Q_1,Q_2,...,Q_4} \psi^{*}(Q_1)\psi(Q_2)\psi^{*}(Q_3)\psi(Q_4)\delta(-Q_1+Q_2-Q_3+Q_4),
\end{split}
\eeq
where $h$ denotes the momentum-independent Yukawa-like coupling introduced above and $\lambda$ stands for the contact four-boson 1PI vertex, which is regenerated during the flow through the box diagram. The non-trivial flow equations for the couplings of the average action (\ref{sb3}) are depicted in Fig. \ref{feyn1} in terms of Feynman diagrams.

It is desirable and possible to cancel the RG flow of the coupling $\lambda$ by absorbing it into the flows of the other couplings. This can be achieved by a rebosonization procedure \cite{Gies, Pawlowski}. The idea is to make the composite field scale-dependent $\phi\to\phi_{k}$ with the choice
\beq \label{sb4}
\partial_{k}\phi_{k}=\alpha_{k}\psi\psi \qquad \partial_{k}\phi^{*}_{k}=\alpha_{k}\psi^{*}\psi^{*},
\eeq
where $\alpha_{k}$ is some real, scale-dependent function. The one-loop flow equation (\ref{fe}) for the average action generalizes to
\beq \label{sb5}
\partial_k \Gamma_k = \frac{1}{2} \text{Tr} \,
  \partial_k R_k \, (\Gamma^{(2)}_k + R_k)^{-1}+\int \frac{\delta \Gamma_k}{\delta \phi_k}\partial_k \phi_{k}+\int \frac{\delta \Gamma_k}{\delta \phi_k^{*}}\partial_k \phi_{k}^{*}.
\eeq
The unknown scale-dependent coefficient $\alpha_k$ can now be fixed by the condition that the flow of the contact coupling $\lambda$ is zero
\beq \label{sb5a}
\partial_t \lambda=0 \Rightarrow \lambda=0.
\eeq
The resulting flow equations (with unspecified cut-off) for the Yukawa coupling $h$ and the inverse propagator $P_{\phi}(Q)$ now read
\beq \label{sb6}
\begin{split}
&\partial_t P_{\phi}(Q)=-\frac{1}{2}h^2 J_{0,d}(Q) \\
&\partial_t h=2h\kappa J_{1,d}+\frac{4\kappa^2 P_{\phi}(Q=0)J_{2,d}}{h},
\end{split}
\eeq
where $J_{n,d}$ was defined in Eq. (\ref{fe7}), and for $J_{0,d}$ we introduce the momentum-dependent generalization
\beq \label{sb7}
J_{0,d}(Q)=\int \frac{d\omega}{2\pi}\frac{d^{d}l}{(2\pi)^d}\partial_t \frac{1}{ [ i(\omega+\omega_{q})+(\vec{l}+\vec{q})^2+R_k(L+Q)][-i\omega+\vec{l}^2+R_k(-L)]}.
\eeq
In order to proceed we must specify the cut-off function and the ansatz for the inverse propagator of the field $\phi$. We choose the sharp cut-off, introduced in Sec. \ref{flowequation} and take the simple ansatz $P_{\phi}(Q)=A_{\phi}(i\omega+\frac{q^2}{2})+m^2$ as used also for the related vacuum 3-body problem in \cite{FSMW}. With this choice the flow equations become
\beq \label{sb8}
\begin{split}
&\partial_t m_{R}^2=(2\xi-d-2)m_{R}^2+\frac{1}{2}h_{R}^2, \\
&\partial_t h_R=(\xi-2-\frac{\kappa}{d-2})h_R-\frac{\kappa^2}{(d-2)^2}\frac{m_{R}^2}{h_R}, \\
&\partial_t A_{\phi R}=(2\xi-d)A_{\phi R}-\frac{1}{4}h_{R}^{2},
\end{split}
\eeq
where we expressed the flows in terms of the rescaled parameters $m_{R}^2=k^{-2-d+2\xi}m^2$, $A_{\phi R}=k^{-d+2\xi} A_{\phi}$, and $h_R=\sqrt{\frac{1}{(4\pi)^{d/2}\Gamma(\frac{d}{2})}}k^{\xi-2}h$. The parameter $\xi$ is not specified yet, but can be identified with the naive scaling dimension of the composite field, i.e. $[\phi]=\xi$. At this point it is convenient to multiply the second equation in (\ref{sb8}) by $2h_{R}$ and obtain flow equations for $m_{R}^2$ and $h_{R}^2$
\beq \label{sb9}
\begin{split}
& \partial_t m_{R}^2=(2\xi-d-2)m_{R}^2+\frac{1}{2}h_{R}^2, \\
& \partial_t h_{R}^2=-\frac{2\kappa^2}{(d-2)^2}m_{R}^2+2(\xi-2-\frac{\kappa}{d-2})h_{R}^{2}.
\end{split}
\eeq
The equations in (\ref{sb9}) form a closed linear system of first order ordinary differential equations. The eigenvalues $\lambda_{\pm}$ of this system can be readily found
\beq \label{sb9a}
\lambda_{\pm}=2\xi-3-\frac{d}{2}-\frac{\kappa}{d-2}\pm\frac{1}{2}\sqrt{(d-2)^2-4\kappa}.
\eeq
We note that $\kappa_{cr}=\frac{(d-2)^2}{4}$, found in Sec. \ref{flowequation}, provides an important boundary value also in the current analysis. While for $\kappa<\kappa_{cr}$ (repulsion and weak attraction) the eigenvalues are purely real, for $\kappa>\kappa_{cr}$ (strong attraction) $\lambda^{\pm}$ acquires a non-trivial imaginary part. 

Let us investigate the latter case in more detail. It is convenient to cancel the real part of  $\lambda^{\pm}$ by choosing $\xi=\frac{1}{2}\left(3+\frac{d}{2}+\frac{\kappa}{d-2} \right)$. With this choice the eigenvalues $\lambda^{\pm}$ are imaginary, leading to a pure oscillatory behavior of the flows of $m_{R}^{2}$ and $h_{R}^2$. Choosing the initial conditions $m_{R}^{2}(t=0)=M$, $h_{R}^{2}(t=0)=H$, we obtain
\beq \label{sb9b}
\begin{split}
& m_{R}^{2}(t)=M \cos (\frac{\sqrt{-D}}{2} t)+W_M \sin (\frac{\sqrt{-D}}{2} t)\\
& h_{R}^{2}(t)=H \cos (\frac{\sqrt{-D}}{2} t)+W_H \sin (\frac{\sqrt{-D}}{2} t),
\end{split}
\eeq
where $D=(d-2)^2-4\kappa$ is the discriminant of the quadratic $\beta$-function introduced in Sec. \ref{flowequation} and $W_M$ and $W_H$ can be expressed as
\beq \label{sb9c}
\begin{split}
& W_M=\frac{1}{\sqrt{-D}}\left(\left[2-d+\frac{2\kappa}{d-2} \right] M+H \right)\\
& W_H=-\frac{1}{\sqrt{-D}}\left(\frac{4\kappa^2}{(d-2)^{2}} M+\left[2-d+\frac{2\kappa}{d-2} \right]H \right).
\end{split}
\eeq
As expected, we obtain a finite (bounded) limit cycle solution (\ref{sb9b}) for the pair $(m_{R}^2, h_{R}^{2})$ for strong attraction $\kappa>\kappa_{cr}$. It is clear that the infinite number of the two-body bound states, present in this case, manifest themselves as zeros of the mass coupling $m_{R}^{2}(t)$ during the RG flow.

\begin{figure} 
	\centering
	\includegraphics[height=1.2in]{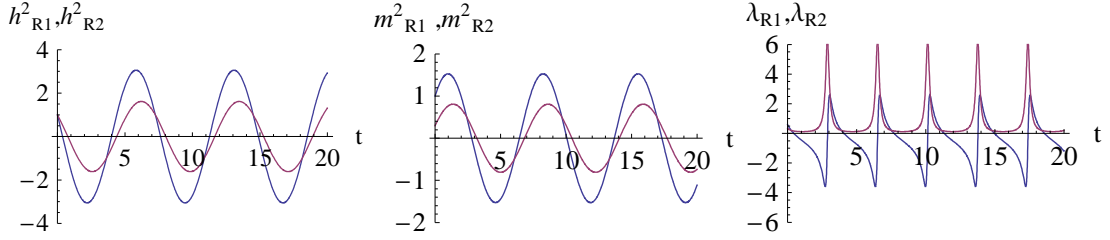}
	\caption{Periodic flows of the real part (blue) and imaginary part (red) of the complex couplings $h_R^{2}$, $m_{R}^2$ and $\lambda_R=\frac{h_{R}^2}{2m_{R}^2}$ for the specific choice of parameters $d=3$ and $\kappa=1$. The amplitude of the $m_{R2}^{2}$ oscillations is related to the decay rate, which must be fixed by experiment. The width of the peaks of $\lambda_{R2}$ is of the order of magnitude of the decay rate.}
\label{bos}
\end{figure}

It is possible to complexify the flow equations (\ref{sb9}) in a similar manner as presented in Sec. \ref{extension} by extending the couplings $h_{R}^{2}$ and $m_{R}^2$ to the complex plane
\beq \label{sb10}
h_{R}^{2}= h_{R1}^{2}+i h_{R2}^{2} \qquad m^{2}_{R}= m^{2}_{R1}+i m^{2}_{R2}.
\eeq
In this extension the non-zero value of $m_{R2}^2$ makes the composite particle $\phi$ unstable. If small compared to $m_{R1}^2$, it equals to the decay width, which is inverse proportional to the life-time of the particle. Due to the linearity of the evolution equations (\ref{sb9}), the flow equations for the pairs $(m_{R1}^{2}, h_{R1}^2)$, $(m_{R2}^{2}, h_{R2}^2)$ decouple. They have exactly the same form as Eq. (\ref{sb9}) with the solution (\ref{sb9b}, \ref{sb9c}), provided we perform the substitution $m_{R}^2\to m_{Ri}^2$, $h_{R}^2\to h_{Ri}^2, M\to M_i, H\to H_i$ with $i=1,2$. We depict the numerical solution of the complexified system in Fig. \ref{bos}. As expected, the flow of the coupling $\lambda_R=\frac{h_{R}^2}{2m_{R}^2}$ reproduces our finding from Fig. \ref{fig1}.

Most importantly, we are now in the position to find an analytical solution of the general non-linear system of flow equations introduced in Sec. \ref{extension}
\beq \label{sb11}
\begin{split}
& \partial_t \lambda_{R1}=\alpha \lambda_{R1}^2-\alpha\lambda_{R2}^2+\beta \lambda_{R1} +\gamma \\
& \partial_t \lambda_{R2}=2\alpha \lambda_{R1}\lambda_{R2}+\beta \lambda_{R2},
\end{split}
\eeq
with $\alpha,\beta,\gamma\in \mathbb{R}$ and $D=\beta^2-4\alpha\gamma<0$. In order to achieve this goal, it is sufficient to project the complex coupling
\beq \label{sb12}
\lambda_{R}=\frac{h_{R}^2}{2 m_{R}^2}=\frac{h_{R1}^{2}+i h_{R2}^{2}}{2[m_{R1}^{2}+i m_{R2}^{2}]}
\eeq
onto its real and imaginary part $\lambda_{R1}$ and $\lambda_{R2}$ 
\beq \label{sb13}
\lambda_{R1}=\frac{m_{R1}^{2} h_{R1}^{2}+m_{R2}^{2} h_{R2}^{2}}{2[m_{R1}^{4}+m_{R2}^{4}]} \qquad \lambda_{R2}=\frac{m_{R1}^{2} h_{R2}^{2}-m_{R2}^{2} h_{R1}^{2}}{2[m_{R1}^{4}+m_{R2}^{4}]}.
\eeq
This expression comprises an explicit solution of the system (\ref{sb11}) provided we substitute
\beq \label{sb14}
\begin{split}
& m_{Ri}^{2}=M_i \cos (\frac{\sqrt{-D}}{2} t)+W_{Mi} \sin (\frac{\sqrt{-D}}{2} t)\\
& h_{Ri}^{2}=H_i \cos (\frac{\sqrt{-D}}{2} t)+W_{Hi} \sin (\frac{\sqrt{-D}}{2} t)
\end{split}
\eeq
with $D=\beta^2-4\alpha \gamma$ and
\beq \label{sb15}
\begin{split}
& W_{Mi}=-\frac{1}{\sqrt{-D}}\left(\beta M_i+ \alpha H_i \right) \\
& W_{Hi}=\frac{1}{\sqrt{-D}}\left(4 \gamma M_i+\beta H_i \right),
\end{split}
\eeq 
where $i=1,2$.

We comment on the issue of initial conditions. Obviously, the first-order system (\ref{sb11}) must be supplemented with two initial conditions $\lambda_{R1}(t=0)$ and $\lambda_{R2}(t=0)$. In our description, however, there are four unknown constants $M_1$, $M_2$, $H_1$ and $H_2$ to be fixed. The identification
\beq \label{sb16}
\lambda_{R1}(t=0)=\frac{1}{2}\frac{M_1 H_1+M_2 H_2}{M_{1}^2+M_{2}^2} \qquad \lambda_{R2}(t=0)=\frac{1}{2}\frac{M_1 H_2-M_2 H_1}{M_{1}^2+M_{2}^2}
\eeq
fixes only two of them, leaving the remaining two ambiguous. This is the known arbitrariness arising in the Hubbard-Stratonovich transformation. Remarkably, one can fix the remaining two constants almost arbitrarily\footnote{Some discrete values of two constants lead to a degenerate solution and are therefore forbidden.}. An especially convenient choice corresponds to $M_1=1$ and $M_2=0$. With this choice we have
\beq \label{sb17}
\lambda_{R1}(t=0)=\frac{1}{2}H_1 \qquad \lambda_{R2}(t=0)=\frac{1}{2}H_2.
\eeq
We present also the analytic solution of Eq. (\ref{sb11}) in the conformal phase with $D>0$. The solution is still given by Eq. (\ref{sb13}) provided we substitute
\beq \label{sb18}
\begin{split}
& m_{Ri}^{2}=M_i \cosh (\frac{\sqrt{D}}{2} t)+W_{Mi} \sinh (\frac{\sqrt{D}}{2} t)\\
& h_{Ri}^{2}=H_i \cosh (\frac{\sqrt{D}}{2} t)+W_{Hi} \sinh (\frac{\sqrt{D}}{2} t)
\end{split}
\eeq
with
\beq \label{sb19}
\begin{split}
& W_{Mi}=-\frac{1}{\sqrt{D}}\left(\beta M_i+ \alpha H_i \right) \\
& W_{Hi}=\frac{1}{\sqrt{D}}\left(4 \gamma M_i+\beta H_i \right).
\end{split}
\eeq 
Finally we note that the analytic solution (\ref{sb12}) can be expressed in the unified, compact form
\beq\label{sb20}
\lambda_R=\frac{1}{2\alpha}\left(-\beta-\sqrt{D}\frac{e^{\frac{\sqrt{D}t}{2}}-C e^{-\frac{\sqrt{D}t}{2}}}{e^{\frac{\sqrt{D}t}{2}}+C e^{-\frac{\sqrt{D}t}{2}}}\right)
\eeq
for both conformal and non-conformal phase. In the last expression $\sqrt{D}\equiv\sqrt{D+i\epsilon}=i\sqrt{-D}$ for $D<0$, and the complex constant $C$ determines the initial condition for $\lambda_R$. Using Eq. (\ref{sb20}) we compute the curvature $\widetilde{\kappa}$ and the related radius $R$ of the complex RG trajectories, which are given by
\beq \label{sb21}
\widetilde{\kappa}=\frac{1}{R}=\frac{|\lambda^{\prime}_{R1}\lambda^{\prime\prime}_{R2}-\lambda^{\prime}_{R2}\lambda^{\prime\prime}_{R1}|}{[\lambda^{\prime 2}_{R1}+\lambda^{\prime 2}_{R2}]^{3/2}}.
\eeq
Here the primes denote the first and second derivative with respect to $t$. 
We observe that the curvature does not depend on the ``RG time'' $t$ in both the conformal and non-conformal phase. For $D>0$ the trajectories form arcs of circles of radius $R=|\frac{C}{2\alpha \text{Im} C}|\sqrt{D}$ (see Fig. \ref{fig2}), while for $D<0$ the trajectories constitute closed circles of radius $R=|\frac{C}{\alpha(1-|C|^2)}|\sqrt{-D}$ (see Fig. \ref{fig1}).

\section{Conclusions} \label{conclusion}
In this work we investigated the nonrelativistic, quantum-mechanical problem of an inverse square potential problem using functional renormalization. The potential is classically scale-invariant. Additionally, it is singular at the origin and the model must be augmented by a contact term, which is necessary for renormalization. We demonstrated that the RG flow of the contact coupling $\lambda$ either approaches a real fixed point (conformal phase) in the IR and UV or undergoes a scale anomaly which manifests itself in an infinite, unbounded limit cycle (non-conformal case). The overall behavior is determined by the sign of the discriminant $D$ of the quadratic $\beta$-function of the contact coupling, and depends on the strength of the long-range inverse square potential and spatial dimension. Remarkably, in the non-conformal phase ($D<0$) the $\beta$-function possesses a pair of complex conjugate fixed points. This observation led us naturally to the extension of the RG analysis to complex values of the coupling $\lambda\to\lambda_1+i\lambda_2$. The RG evolution was computed numerically and the phase portraits of the flows were obtained. Additionally, we provided a geometric description of the complex flows on the Riemann sphere. We observed that in the non-conformal phase the real part $\lambda_1$ and the imaginary part $\lambda_2$ develop a finite (bounded) limit cycle in the complex plane. This suggests that the complex extension can be utilized as an efficient numerical tool for investigation of various problems with infinite (unbounded) limit cycles, as for example, in few-body physics of cold atoms near unitarity. One should simply add a small imaginary part $\lambda_2$ to the real part $\lambda_1$ and follow the extended RG evolution. This procedure regulates the periodic singularities during the RG flow making the numerical treatment feasible. From a physical point of view, the introduction of the complex coupling constant appears quite natural since it arises from an inelastic channel in the two-body scattering. The bosonization procedure allowed us to view the problem from a different perspective. More importantly, it enabled us to find an analytical solution of the extended set of non-linear flow equations.

Quite surprisingly, a similar RG behavior as in our case of an inverse square potential appears also in the context of chiral dynamics in QCD with a large number of flavors \cite{Gies06,Braun07}. Here, gluon exchange between the quarks induces quark self-interactions. The flow equations governing the RG evolution of these self-interactions exhibit the same behavior as shown in Fig. \ref{figure2}. Depending on the value of the gauge coupling $\alpha$ the flow equations possess either two fixed points ($\alpha<\alpha_{cr}$) and the system is in the chiral symmetric phase, or as $\alpha$ increases above this critical value $\alpha_{cr}$ one ends up in the chiral symmetry broken phase.  The analog to the conformal symmetry breaking in dependence on the value of $\kappa$ in our simple case of an inverse square potential appears to be quite remarkable.

Finally, we comment on a possible connection between our work and the recent studies of non-Fermi liquids using the AdS/CFT correspondence \cite{Lee,Liu,Faulkner}. The authors of \cite{Liu,Faulkner} studied the relativistic many-body physics of fermions at vanishing temperature in $d$ spacetime dimensions by mapping onto a classical gravity problem with an extremal charged black hole in anti-de Sitter spacetime ($AdS_{d+1}$). In this description the low-energy scaling behavior around the Fermi surface is related to the near-horizon geometry, which turns out to be $AdS_{2}\times \mathbb{R}^{d-1}$. Notably, the isometry group of the $AdS_{2}$ part is $SO(2,1)$, which is exactly the symmetry group of the quantum mechanics of the inverse square potential in the conformal phase \cite{Camblong}. This suggests that the emergent IR CFT, defined in \cite{Faulkner}, might be conformal quantum mechanics with the inverse square potential. Another evidence in this direction is given by the observation that the low-energy behavior of the real and imaginary parts of the retarded Green functions, computed numerically in \cite{Liu, Faulkner}, agrees remarkably well with the RG flows of the real and imaginary part of the contact coupling in conformal and non-conformal phases (see Figs. \ref{fig1}, \ref{fig2}). This suggests that the complex extension and its analytical solution, found in this work, might be useful for a better understanding of non-fermi liquids from the AdS/CFT correspondence.  

\emph{Acknowledgments} -- It is our pleasure to acknowledge the discussions with S. Floerchinger, H.-W. Hammer, J. M. Pawlowski, D. T. Son, C. Wetterich and W. Zwerger. We are thankful to J. Ho\v{s}ek for critical reading of the manuscript. SM is grateful to KTF for support. RS thanks the DFG for support within the FOR 801 'Strong correlations in multiflavor ultracold quantum gases'.

\appendix
\section{Complex RG flows on the Riemann sphere} \label{appendix}
\begin{figure}
     \centering
     \includegraphics[height=5in]{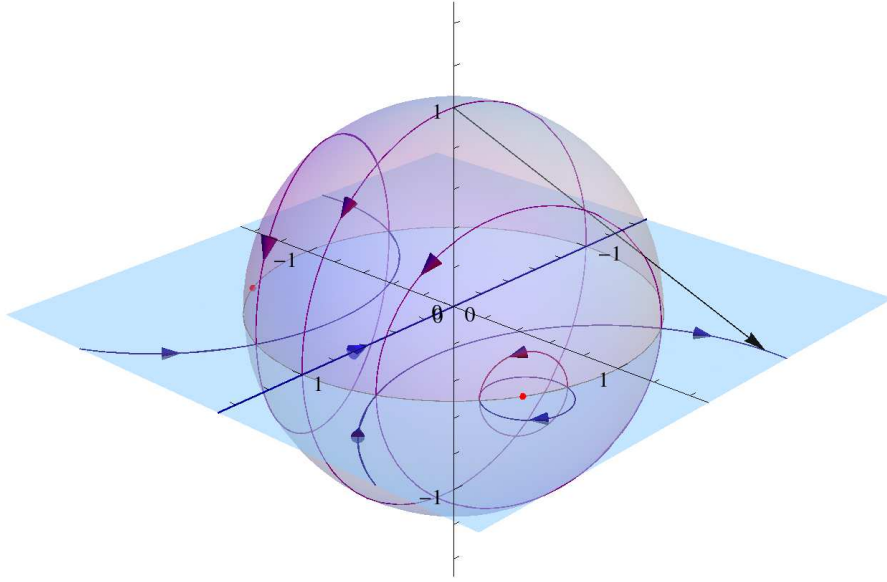}
    \vspace{-4cm}
     \caption{RG flows on the complex plane and the corresponding map onto the Riemann sphere in the non-conformal phase. The pink dots represent the two complex fixed points. The solid lines with arrows, pointing towards the UV, correspond to different RG trajectories. Additionally, the black straight solid line from the north pole illustrates the stereographic projection of the Riemann sphere $S^2$ onto the complex plane $\mathbb{C}$.}
 \label{riemann-}
 \end{figure}

\begin{figure}
     \centering
     \includegraphics[height=5in]{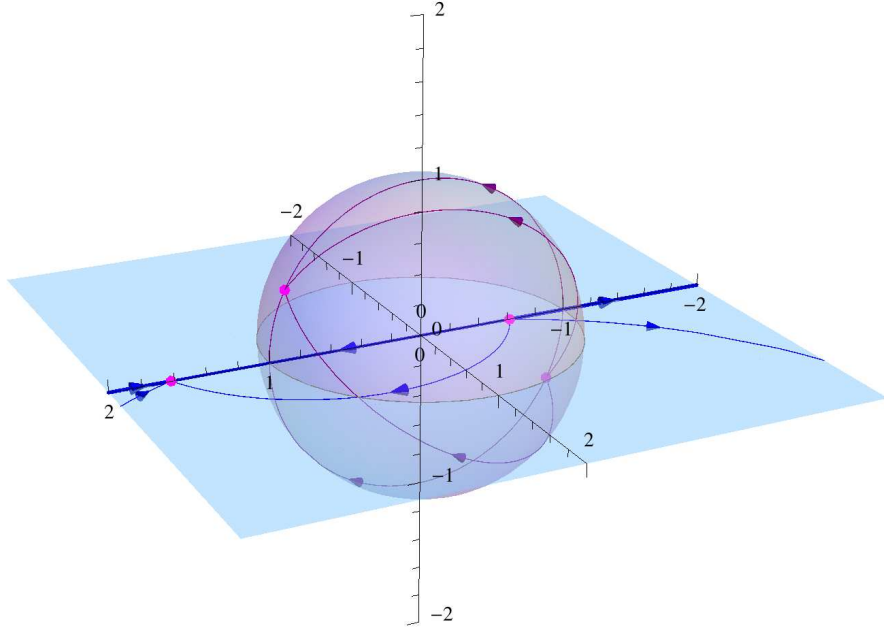}
    \vspace{-4cm}
     \caption{RG flows on the complex plane and the corresponding map onto the Riemann sphere in the conformal phase. The pink dots represent the two real fixed points. The solid lines with arrows, pointing towards the UV, correspond to different RG trajectories.}
 \label{riemann+}
 \end{figure}

As has already been noted in Sec. \ref{extension}, the generalized RG flow of the complex coupling $\lambda=\lambda_1+i\lambda_2$ can be most conveniently studied on the Riemann sphere. The Riemann sphere is a unit sphere $S^{2}$ in three dimensional real space $\mathbb{R}^3$, which intersects the complex plane $\overline{\mathbb{C}}=\mathbb{C}\cup\{\infty\}$ at the equator. The map $\sigma \text{:} \hspace{2mm} S^{2}\to\overline{\mathbb{C}}$ onto the complex plane $(\lambda_1,\lambda_2)$ is given by the stereographic projection (see Fig. \ref{riemann-}) and reads in the Cartesian coordinates $(x, y, z)\in S^{2}$
\begin{eqnarray}
\lambda_1=\frac{x}{1-z},\quad\lambda_2=\frac{y}{1-z}\qquad \textrm{if}\quad(x,y,z)\neq(0,0,1)\notag\\
(\lambda_1,\lambda_2)=\infty\qquad \textrm{if}\quad (x,y,z)=(0,0,1).
\end{eqnarray}
The inverse map is given by
\begin{eqnarray} \label{a2}
x=\frac{2\lambda_1}{\lambda_1^2+\lambda_2^2+1}, \qquad y=\frac{2\lambda_2}{\lambda_1^2+\lambda_2^2+1} ,\qquad z=\frac{\lambda_1^2+\lambda_2^2-1}{\lambda_1^2+\lambda_2^2+1},\qquad \lambda\neq\infty\notag\\
x=0,\qquad y=0,\qquad z=1,\qquad\lambda=\infty.
\end{eqnarray}
The analytical solution of the complex flow equations (\ref{c2}) was obtained in Sec. \ref{bosonization}. We substitute these solutions into Eq. (\ref{a2}) and obtain the analytical solution on the Riemann sphere, which we plot for both the non-conformal ($D<0$) and the conformal ($D>0$) phases in Figs. \ref{riemann-}, and \ref{riemann+}.

\clearpage

\end{document}